\pdfoutput=1
\documentclass[journal,twoside,web]{ieeecolor}
\usepackage{tmi}
\usepackage{cite}
\usepackage{amsmath,amssymb,amsfonts}
\usepackage{algorithmic}
\usepackage{graphicx}
\usepackage{textcomp}
\def\BibTeX{{\rm B\kern-.05em{\sc i\kern-.025em b}\kern-.08em
    T\kern-.1667em\lower.7ex\hbox{E}\kern-.125emX}}

\usepackage{url}

\usepackage{bbm}
\usepackage{ctable}
\usepackage{multirow}
\usepackage[font={footnotesize,sf}]{subcaption}
\usepackage[font={footnotesize,sf}]{caption}
\usepackage{cleveref}
    
\graphicspath{{figures/}}
\DeclareGraphicsExtensions{.pdf,.png}

\newcommand{\alignedintertext}[1]{%
  \noalign{%
    \vskip\belowdisplayshortskip
    \vtop{\hsize=\linewidth#1\par
    \expandafter}%
    \expandafter\prevdepth\the\prevdepth
  }%
}

\newcommand{\mcc}[1]{\multicolumn{1}{c}{#1}} 
\newcommand{\etal}{\mbox{\emph{et al.}}}

\begin{document}
\bstctlcite{IEEEexample:BSTcontrol}
\title{Echocardiography Segmentation with Enforced Temporal Consistency}
\author{Nathan Painchaud, Nicolas Duchateau, Olivier Bernard, and Pierre-Marc Jodoin
\thanks{Manuscript received November 26, 2021; revised April 20, 2022. This work was supported in part by the NSERC Discovery Grants' program, the NSERC Canada Graduate Scholarships-Doctoral program, the FRQNT Doctoral Scholarships program and the LABEX PRIMES (ANR-11-LABX-0063) of Université de Lyon, within the program "Investissements d'Avenir" (ANR-11-IDEX-0007) operated by the French National Research Agency (ANR). }
\thanks{N. Painchaud and P.-M. Jodoin are with the Department of Computer Science, University of Sherbrooke, Sherbrooke, QC, Canada (e-mail: nathan.painchaud@usherbrooke.ca). }
\thanks{N. Painchaud, N. Duchateau, and O. Bernard are with Univ Lyon, INSA‐Lyon, Université Claude Bernard Lyon 1, UJM-Saint Etienne, CNRS, Inserm, CREATIS UMR 5220, U1294, F‐69621, Lyon, France. }
\thanks{N. Duchateau is also with the Institut Universitaire de France (IUF). }
\thanks{Copyright \textcopyright 2022 IEEE. Personal use of this material is permitted. However, permission to use this material for any other purposes must be obtained from the IEEE by sending a request to pubs-permissions@ieee.org.}}

\maketitle

\begin{abstract}
Convolutional neural networks (CNN) have demonstrated their ability to segment 2D cardiac ultrasound images. However, despite recent successes according to which the intra-observer variability on end-diastole and end-systole images has been reached, CNNs still struggle to leverage temporal information to provide accurate and temporally consistent segmentation maps across the whole cycle. Such consistency is required to accurately describe the cardiac function, a necessary step in diagnosing many cardiovascular diseases. In this paper, we propose a framework to learn the 2D+time apical long-axis cardiac shape such that the segmented sequences can benefit from temporal and anatomical consistency constraints. Our method is a post-processing that takes as input segmented echocardiographic sequences produced by any state-of-the-art method and processes it in two steps to (i) identify spatio-temporal inconsistencies according to the overall dynamics of the cardiac sequence and (ii) correct the inconsistencies. The identification and correction of cardiac inconsistencies relies on a constrained autoencoder trained to learn a physiologically interpretable embedding of cardiac shapes, where we can both detect and fix anomalies. We tested our framework on 98 full-cycle sequences from the CAMUS dataset, which are available alongside this paper. Our temporal regularization method not only improves the accuracy of the segmentation across the whole sequences, but also enforces temporal and anatomical consistency.
\end{abstract}

\begin{IEEEkeywords}
Deep learning, CNN, variational autoencoder, cardiac segmentation, ultrasound, left ventricle, myocardium.  \vspace{-0.5cm}
\end{IEEEkeywords}

\section{Introduction}
\label{sec:introduction}
\IEEEPARstart{I}{n} clinical practice, ultrasound (US) imagery is the modality of choice to routinely evaluate the cardiac function. Its popularity derives from  the fact that it is real-time, low-cost and noninvasive. However, these advantages come at the cost of low quality images in comparison to other modalities like CT scans and magnetic resonance imaging (MRI). For that reason, automatic methods have always struggled to properly analyse US data. Nevertheless, in the past few years, multiple independent studies have shown how accurate well-trained CNNs can be at segmenting the main sections of the heart, namely the ventricles, the atria and the myocardium. The most effective neural networks can even reach the intra-observer variability~\cite{wei_temporal-consistent_2020}.

However, up until now, state-of-the-art (SOTA) US segmentation methods have focused on 2D CNNs designed to process key moments of the cardiac cycle, namely the end-diastolic (ED) and end-systolic (ES) instants. This focus on static 2D images neglects the temporal richness of US, which is one of its selling points for day-to-day clinical use. Even methods that produce segmentation maps over the whole cardiac cycle, like the one proposed by Wei \etal~\cite{wei_temporal-consistent_2020} where the temporal aspect adds context to ED and ES segmentation, are mostly validated within the scope of 2D segmentation at the ED and ES instants.

This focus on 2D segmentation can be explained by the fact that the ejection fraction (EF), measured between the ED and ES segmentations, is widely reported as a valuable indicator of various cardiovascular diseases~\cite{mcdonagh_2021_2021}, and thus is commonly used to evaluate the clinical performance of a segmentation method. Unfortunately, the EF discards the intermediate frames between ED and ES, even though they provide useful information to characterize other pathologies~\cite{smiseth_myocardial_2016,cikes_beyond_2016}.

To the best of our knowledge, the machine learning papers that have so far studied the temporal consistency of 2D+time segmentation methods in US have limited themselves to qualitative evaluations based on global indices. Thus, we first present an exhaustive analysis of the shortcomings of SOTA methods when it comes to producing temporally consistent segmentations, and then introduce quantifiable and clinically significant metrics to evaluate temporal consistency. Afterwards, we propose a novel generic temporal regularization procedure that corrects those shortcomings, to improve the overall accuracy of the temporal segmentations and increase their physiological relevance. Our main contributions are:
\begin{enumerate}
    \item We define clinically interpretable indicators to quantitatively evaluate the temporal consistency of 2D+time segmentations;
    \item We introduce a generic post-processing algorithm, based on an interpretable embedding of cardiac shapes, which can be plugged at the end of any segmentation method, which enforces temporal consistency on top of improving the overall accuracy of the segmentations\footnote{Code is available at \href{https://github.com/vitalab/castor}{https://github.com/vitalab/castor}};
    \item We make public a new fully-annotated dataset of 98 full cycle apical 4 chamber (A4C) sequences from the CAMUS dataset. Up until now, only ED and ES expert annotations were available for these US sequences. As far as we know, this is the first public dataset of its kind for 2D echocardiography\footnote{Dataset is available for download at \href{https://humanheart-project.creatis.insa-lyon.fr/ted.html}{https://humanheart-project.creatis.insa-lyon.fr/ted.html}}.
\end{enumerate}

\vspace{-0.08cm}
\section{Previous Works}
\label{sec:previous}
Because our method processes segmentations while relying on a learned embedding of cardiac shapes, we build on top of two separate research domains: the applied domain of echocardiographic segmentation and the more theoretical domain of interpretable representation learning.

\vspace{-0.08cm}
\subsection{Echocardiographic Segmentation}
\label{sec:echo_segmentation}

\subsubsection{2D CNNs}
In 2019, Leclerc \etal~\cite{leclerc_deep_2019} established a milestone in echocardiographic segmentation by publishing the CAMUS dataset, containing 500 patients, distributed across 450 training patients with public expert annotations and 50 testing patients, for which the expert annotations are not publicly available. For each patient in the training dataset, images and expert annotations of the left ventricle (LV), myocardium (MYO) and left atrium (LA) were provided for the ED and ES frames of the cardiac cycle, in both apical 4 chamber (A4C) and apical 2 chamber (A2C) views. In addition to making the CAMUS dataset publicly available, the authors also demonstrated competitive results with different versions of a UNet~\cite{ronneberger_u-net_2015} finetuned to the task.

More recently, Leclerc \etal~introduced LU-Net~\cite{leclerc_lu-net_2020}, a two-step segmentation network inspired by Mask R-CNN where a first network predicts a region of interest (ROI) around the heart and a second network predicts a highly accurate segmentation within the ROI. LU-Net showed improved results compared to the authors' previous UNet architecture, even beating intra-observer accuracy on the epicardium.

Other works proposed to tweak the base UNet architecture to improve its performance for 2D echocardiographic segmentation. For instance, Maraldi \etal~\cite{moradi_mfp-unet_2019}, inspired by feature pyramid networks (FPN), proposed MFP-Unet, which uses dilated convolutions to increase the receptive field and up-scales feature maps to maximize the information available in the last layers.

In parallel to the work by Leclerc \etal, Ouyang \etal~published their own dataset, called EchoNet-Dynamic~\cite{ouyang_echonet-dynamic_2019}.  This dataset focuses on evaluating the ejection fraction (EF) and segmenting the LV in A4C sequences. In a follow-up paper studying cardiac function using the EchoNet dataset, Ouyang \etal~\cite{ouyang_video-based_2020} relied on the generic DeepLabv3 architecture~\cite{chen_encoder-decoder_2018} for their segmentation network.

Independently from Ouyang \etal, Painchaud \etal~\cite{painchaud_cardiac_2020} further demonstrated, in a general study on cardiac segmentation spanning MRI and US images, that generic networks such as ENet~\cite{paszke_enet_2016} can perform competitively with respect to networks optimized with echocardiography segmentation in mind, like the UNet and LU-Net from Leclerc \etal

\subsubsection{2D+time CNNs}
While most research making use of CAMUS and EchoNet-Dynamic introduced variations of generic architectures aimed at 2D segmentation, a few papers proposed specialized frameworks to focus on the temporal aspect of echocardiography.

PV-LVNet~\cite{ge_pv-lvnet_2019}, proposed by Ge \etal, does not provide intermediate segmentations, but rather predicts multiple indices related to the cardiac function. Leveraging temporal information with a recurrent network, PV-LVNet aims to reliably locate and crop the LV across whole US sequences. From these cropped sequences, another recurrent network predicts more accurate indices, not unlike LU-Net did for segmentation. Finally, PV-LVNet also makes use of multi-view information, providing cropped sequences from both A2C and A4C sequences to a single network to predict the LV volume.

Shortly after, MV-RAN~\cite{li_mv-ran_2020} tried like PV-LVNet to exploit both temporal and multi-view information, this time to perform segmentation of the LV. Similar to MFP-Unet, MV-RAN uses dilated convolutions in the encoder to extract multi-scale features. The stack of 2D features from a sequence are then processed by two collaborating branches: a 3D CNN that classifies the sequence's view, and a convolutional LSTM that outputs the 2D+time segmentation. The 2D segmentation performance were reported for the ED and ES frames of the CAMUS dataset, but the results relating to temporal consistency were only provided on a private dataset.

Most recently, Wei \etal~introduced CLAS~\cite{wei_temporal-consistent_2020}, a novel 3D segmentation network that strives to be temporally consistent while relying only on ED and ES annotations. To achieve this, CLAS  predicts deformation fields and uses it to propagate the annotations during training. Wei \etal~showed that their method improves the consistency between ED and ES predictions, landing within the intra-observer variability. However, their analysis of the temporal consistency across the sequences was limited to qualitative evaluations over a few patients.

\begin{figure*}[tp]
    \includegraphics[width=\textwidth]{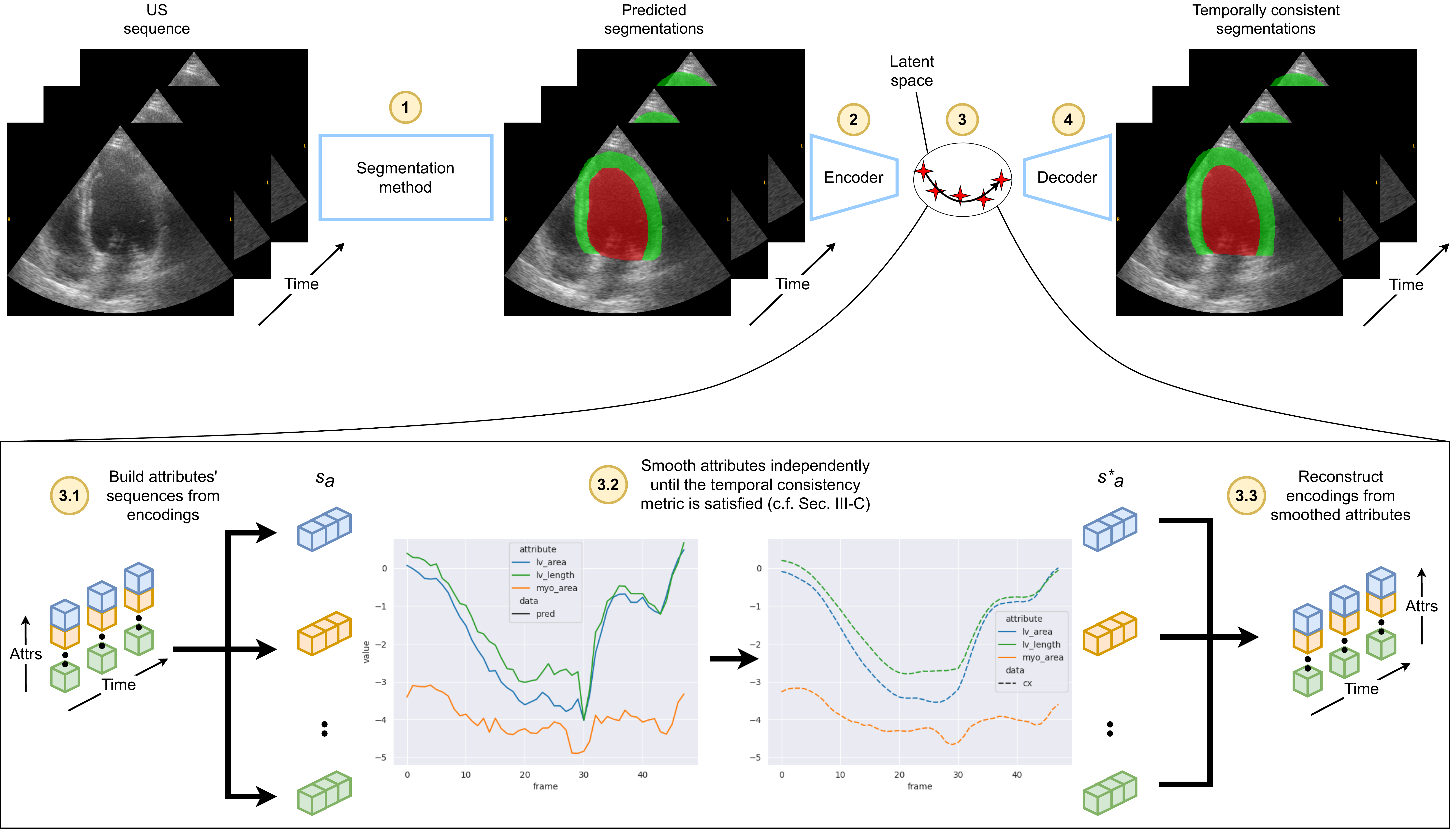}
    \caption{Schematic representation of our temporal regularization method. Starting with the raw echocardiography sequence, a SOTA segmentation method predicts a segmentation mask (1), which is then encoded frame-by-frame (2) by a pretrained autoencoder. There, the encodings of the sequence are split between dimensions (3.1) to produce sequences of attributes w.r.t time ($s_a$). These sequences are then processed invidually (3.2), and the results ($s^*_a$) are merged back together as encodings for each frame (3.3). Finally, the modified encodings are decoded into now temporally consistent segmentations (4). \vspace{-0.3cm}}
    \label{fig:method}
\end{figure*}

\vspace{-0.08cm}
\subsection{Interpretable Representation Learning}
\label{sec:representation_learning}

\subsubsection{In computer vision} The learning of interpretable representations has been tied to the learning of disentangled representations, where the underlying \emph{data generative factors}, that are independent from one another, have to be aligned with specific dimensions of the embedding. The disentangled data generative factors learned by models correspond, more often than not, to human-interpretable, high-level semantic concepts present in the images~\cite{higgins_beta-vae_2017,burgess_understanding_2018}.

However, much of the research into disentangled representation learning has targeted unsupervised applications on generic datasets, where it is not possible to assume any prior knowledge about possible data generative factors~\cite{tschannen_recent_2018,kim_disentangling_2018,chen_isolating_2018}. While useful to improve our fundamental understanding of representation learning, this bias towards unsupervised disentanglement is less relevant in medical image analysis, where we have a wealth of medical knowledge to guide the learning of clinically relevant factors of variation.

The fader networks from Lample \etal~\cite{lample_fader_2017} constituted a major stepping stone in learning interpretable representation with some supervision. Using a discriminator on supervised labels, the fader networks learns an embedding invariant to these labels. In practice, this allows users to control, at inference time, the presence or absence of specific generative factors, i.e. the labels, provided as complementary input to the encoder's embedding to the decoder. However, when testing fader networks, we encountered two major drawbacks reported in subsequent research~\cite{pati_attribute-based_2021}: they do not scale well to multiple independent generative factors, and they work mostly on categorical factors of variation, not generalizing well to continuous ones.

Recently, the Attribute-Regularized VAE (AR-VAE) introduced by Pati \etal~\cite{pati_attribute-based_2021} remedies the main issues with fader networks and is able to correlate different dimensions of the embedding to different generative factors, which they call \emph{attributes}. The AR-VAE achieves this using a sorting penalty (from multiple data samples) along a dimension in the embedding to be in the same order as an attribute in the images. This notion of order allows the AR-VAE to work well with continuous attributes, and in practice it also scales well to multiple attributes, each being correlated to its own dimension in the embedding.

\subsubsection{In cardiac medical imaging} Disentangled representations played a role in recent works on cardiac MRI segmentation by Chartsias \etal~\cite{chartsias_disentangled_2019} and Valvano \etal~\cite{valvano_temporal_2019}. However, their latent space consists of binary masks at the same resolution as the images, and is therefore too high-dimensional to be used as an embedding in which to easily manipulate samples. Clough \etal~\cite{clough_global_2019} and Puyol-Antón \etal~\cite{puyol-anton_interpretable_2020} proposed an interpretable VAE-based model that characterizes the impact of clinically relevant factors of variation. Still, they do not offer a way to leverage interpretability when generating samples. At the other end of the spectrum, Painchaud \etal~\cite{painchaud_cardiac_2020} used a constrained variational autoencoder to efficiently manipulate samples in a low-dimensional latent space, but their representation overall lacked interpretability. The work that tackled interpretability most like we do came from Zhang \etal~\cite{zhang_orthogonal_2017}, where cardiac shapes are decomposed into orthogonal, i.e. disentangled, components related to clinical indices. However, the input data consists of LV shape models instead of image segmentations, and the authors used linear methods which are suited for such data, but not necessarily for high-dimensional image data.

\section{Proposed Framework}
\label{sec:framework}

We give a schematic representation of our method in \cref{fig:method}. The system first uses an arbitrary SOTA segmentation method to process the US sequence into a sequence of 2D segmentations, which are then fed to a pretrained 2D cardiac shape autoencoder that corrects temporal inconsistencies. The temporal consistency is the result of a \emph{temporal regularization} optimization operation in the latent space of the autoencoder, which reconstructs all the frames in the sequence, targeting the inconsistent frames especially to warp them towards an interpolation of their neighboring frames.

The temporal consistency is determined by looking at the variation over time of key cardiac shape attributes. Since the cardiac motion is locally monotonic, we can expect the cardiac attributes' values with respect to time to have the same properties. Thus, we consider sudden and significant reversals of order in these attributes to be indicative of temporal inconsistencies in the segmentation. Following this, the aim of our system is to output 2D+time segmentations that always respect the locally monotonic nature of the shape attributes over the duration of the whole cardiac cycle.

\subsection{Cardiac Shape Autoencoder}
\label{sec:cardiac_ar-vae}

In the context of implementing the temporal regularization operation, learning an interpretable embedding of 2D cardiac shapes provides an unrivaled advantage: it allows to monitor \emph{temporal consistency indicators}, normally done in image space, in the same space where we can meaningfully manipulate cardiac shapes. It also implicitly models time, as a path between frames in the latent space, without including it as a third dimension of the data, since doing so is known to require considerably more data and to produce images of significantly worse quality~\cite{volokitin_modelling_2020}.

\begin{figure*}[tp]
    \includegraphics[width=\textwidth]{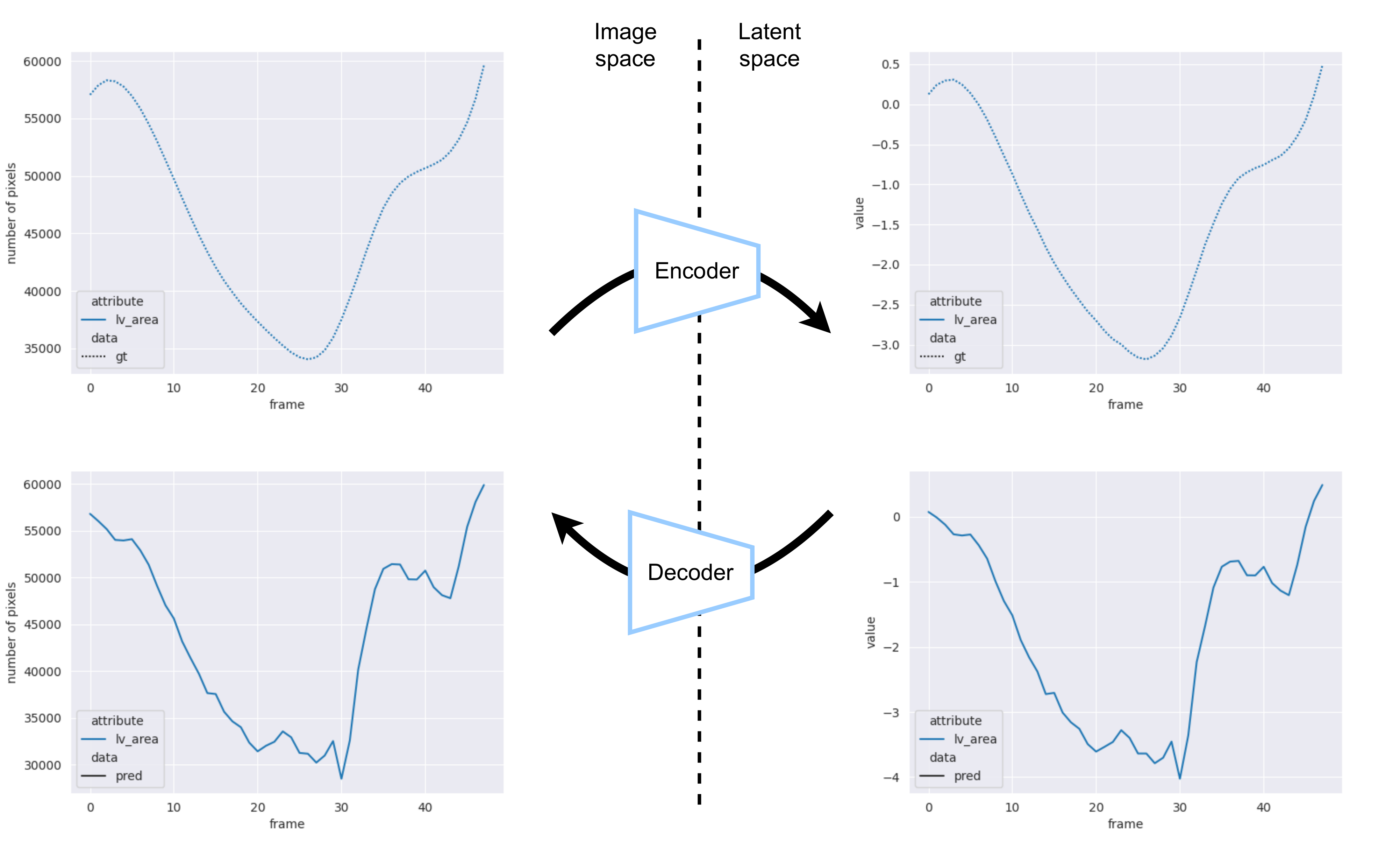}
    \caption{Examples of correlation between image attributes and latent attributes. Plots on the [left] are of attributes w.r.t. time computed on sequences of segmentations, while plots on the [right] are of the correlated dimensions in the encodings of the same sequences. Plots at the [top] are of temporally consistent ground truth (\textit{gt}) segmentations, while the [bottom] plots are of temporally inconsistent predictions (\textit{pred}) made by a UNet.}
    \label{fig:cardiac_attributes_correlation}
\end{figure*}

In practice, we learn the interpretable embedding without requiring any additional human annotations, in a self supervised manner. Based on experts' prior knowledge of cardiac anatomy, we came up with seven clinically relevant attributes characterizing a 2D cardiac shape: i) LV area, ii) LV width at the valves, iii) LV length from valves to apex, iv) LV principal axis orientation, v) MYO area, vi) epicardium (EPI) horizontal center of mass, vii) EPI vertical center of mass. When observing their variations over a cardiac cycle, these attributes all evolve continuously. Thus, we opted to use an AR-VAE (see \cref{sec:representation_learning}) as the framework for our cardiac shape autoencoder, since it is the best suited method we are aware of to handle continuous attributes. We implemented the AR-VAE loss as described in the paper by Pati \etal~\cite{pati_attribute-based_2021}, but otherwise used our own convolutional autoencoder architecture and training hyperparameters, tuned to fit the CAMUS dataset (see \cref{sec:cardiac_ar-vae_details}). Most importantly, we use a latent space with 16 dimensions, setting aside 7 dimensions for our attributes, and leaving 9 \emph{residual dimensions} to encode additional data. \Cref{fig:cardiac_attributes_correlation} gives examples of the correlation between the attributes computed on the segmentations and their corresponding dimensions in the latent space encodings, called \emph{latent attributes}.

\subsubsection*{Cardiac Shape Manipulation}

\begin{figure}[tp]
    \begin{subfigure}[b]{\columnwidth}
        \includegraphics[width=\textwidth]{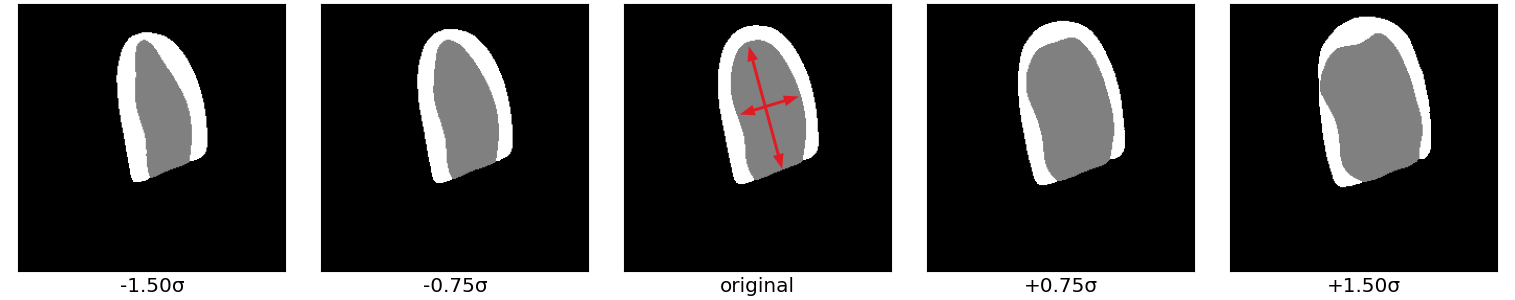}
        \caption{Left ventricle (LV) area}
    \end{subfigure}
    \begin{subfigure}[b]{\columnwidth}
        \includegraphics[width=\textwidth]{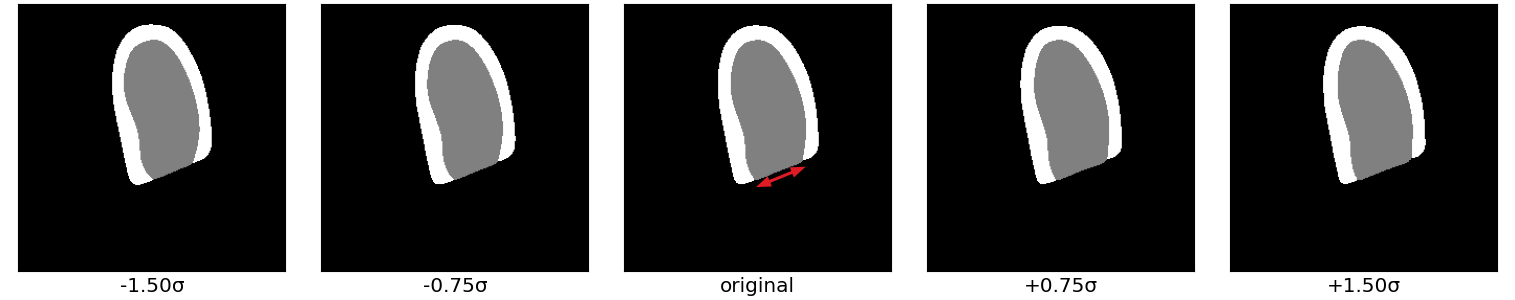}
        \caption{LV width at the valves}
    \end{subfigure}
    \begin{subfigure}[b]{\columnwidth}
        \includegraphics[width=\textwidth]{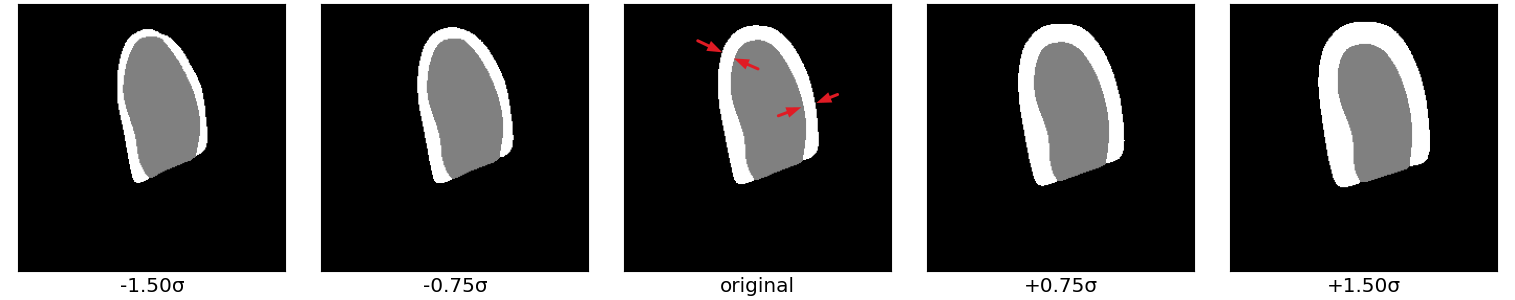}
        \caption{Myocardium (MYO) area}
    \end{subfigure}
    \begin{subfigure}[b]{\columnwidth}
        \includegraphics[width=\textwidth]{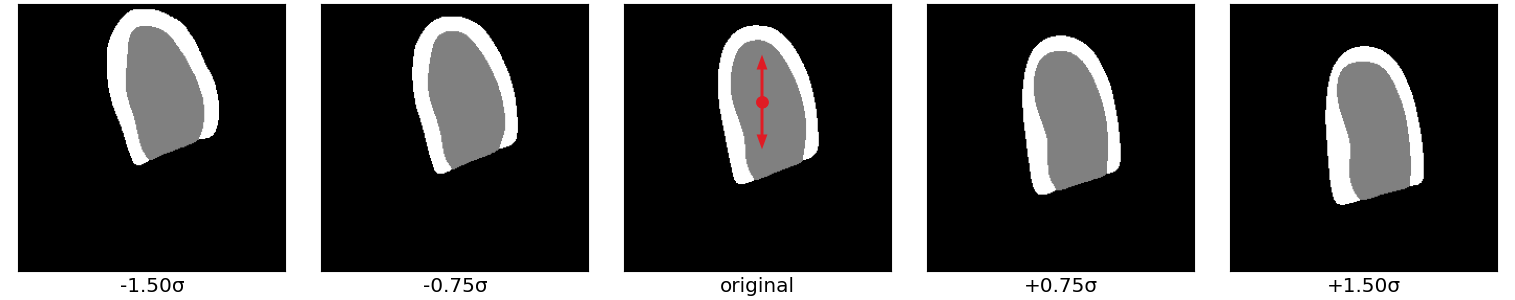}
        \caption{y-coordinate of the cardiac center of mass}
    \end{subfigure}
    \caption{Samples generated by sweeping over different attributes in the latent space (only one at a time) around a sample from the dataset. The labels below each image indicate the value added to the latent dimension, where $\sigma$ is that dimension's standard deviation from the aggregate posterior over the whole dataset.}
    \label{fig:cardiac_shape_manipulation}
\end{figure}

Once fully-trained, seven dimensions in the latent attributes are tailored to the cardiac attributes. The interpretability of the embedding learned by the cardiac AR-VAE allows us to manipulate each of the seven attributes independently, with as little impact on the rest of the cardiac shape as necessary for it to remain consistent and anatomically plausible. \Cref{fig:cardiac_shape_manipulation} is a demonstration of this capability, where we start from a reference image (center), encode it, modify a single interpretable latent dimension, and decode it back to an image.

\subsection{Temporal Consistency Indicator}
\label{sec:temporal_indicator}

It is possible to detect temporal inconsistencies by analyzing the variations over time of the aforementioned attributes, whether they are extracted from the images or the latent space. However, the values have to be normalized to be comparable between domains. Given $A_d$ the set of all values of an attribute in a domain $d$, each temporal sequence $s_a$ of an attribute $a$ is normalized according to the following equation:
\begin{equation}
s_a \leftarrow \frac{s_a - \min(A_d)}{\max(A_d) - \min(A_d)}
\end{equation}

\begin{figure}[tp]
    \begin{subfigure}[b]{\columnwidth}
        \includegraphics[width=\textwidth]{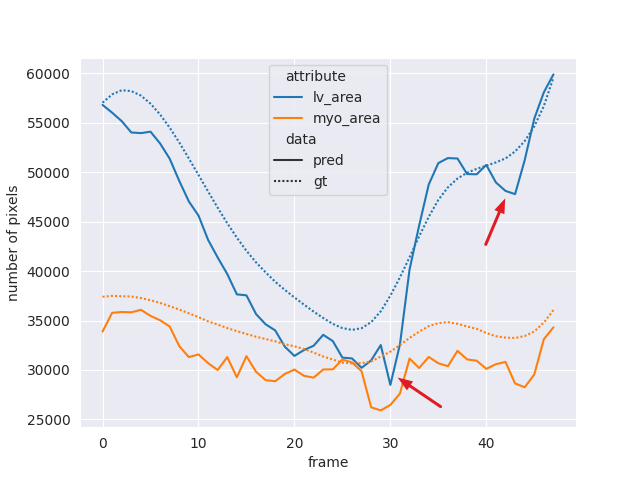}
        \caption{Inconsistent LV area, manifesting in the segmentations as choppy contractions/dilations of the LV}
    \end{subfigure}
    \begin{subfigure}[b]{\columnwidth}
        \includegraphics[width=\textwidth]{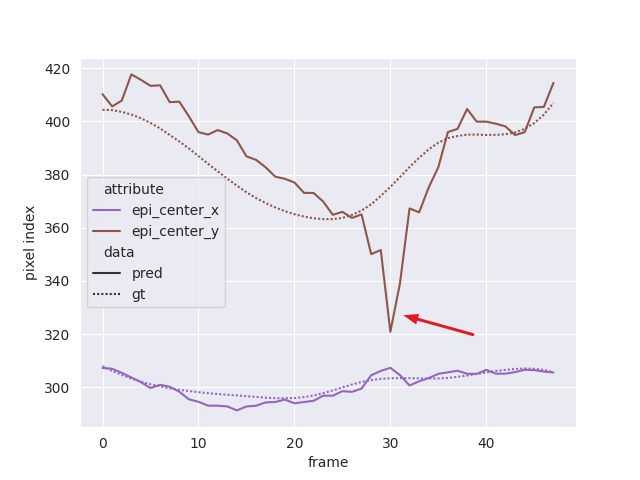}
        \caption{Inconsistent y-coordinate of the EPI's center of mass, manifesting in the segmentations as abrupt vertical shifts of the cardiac shape}
    \end{subfigure}
    \caption{Examples of temporally inconsistent segmentations, considering different cardiac anatomical characteristics over time. The different colors represent different attributes. The smooth dotted lines (\textit{gt}) correspond to attributes of the ground truth segmentations, while the chaotic solid lines (\textit{pred}) represent the attributes extracted from a SOTA segmentation method's predictions. For these plots, the SOTA segmentation method was the UNet from Leclerc \etal~\cite{leclerc_deep_2019}.}
    \label{fig:temporal_inconsistencies_examples}
\end{figure}

After normalization, substantial erratic changes over time of any of the attributes are indicative of inconsistent inter-frame segmentations, i.e. temporal inconsistencies (see the attribute plots in \cref{fig:temporal_inconsistencies_examples}).  One way to measure the temporal smoothness of an attribute is through its second-order derivative: $\frac{d^2 s_{a}(t) }{dt^2}$. As such, a large derivative is a sign of intense variation while a small derivative indicates local smoothness.  One may turn this measure into a discriminative indicator function:
\begin{equation}
\label{eq:temporal_indicator}
\mathbbm{1}(s_a,t) \overset{\mathrm{def}}{=\joinrel=} \left | \frac{d^2 s_{a}(t) }{dt^2} \right | > \tau_a
\end{equation}
which is equal to $1$ when the second-order derivative is above a threshold $\tau_a$ and $0$ otherwise.   Here $\tau_a$ is specific to the attribute $a$  and corresponds to the upper bound of acceptable deviations.

Since cardiac time frames are discrete by nature, one may numerically approximate the second-order derivative as follows:
\begin{equation}
\label{eq:local_linearity}
\frac{d^2 s_{a}(t) }{dt^2} \approx s_{a,t+1} + s_{a,t-1} - 2s_{a,t}
\end{equation}
which corresponds to a Laplacian filter measuring how three consecutive values are temporally aligned along the cardiac cycle. The seven thresholds $\tau_a$ (one for each attribute) are determined empirically from the maximum value observed for each attribute in the training data, to reach high recall for temporal inconsistencies, and then manually raised based on the inspection of the evaluation on segmentation methods to ensure perfect precision for temporal inconsistencies.

To help visualize the kind of patterns that are considered temporally consistent or not, \cref{fig:temporal_inconsistencies_examples} illustrates typical variations in attributes' values with respect to time that get flagged as temporally inconsistent by a Laplacian filter. These plots show local spikes along the temporal curve of the left ventricle and myocardium areas as well as the epicardium center position.  In the next sub-section, we explain how temporal regularization can compensate for these unwanted spikes.

\subsection{Temporal Regularization}
\label{sec:temporal_regularization}

The temporal regularization operation is where the interpretable cardiac shape autoencoder and temporal Laplacian consistency indicator come together. The reader might follow the explanation through the diagram in \cref{fig:method}. Starting with a sequence of 2D segmentation maps, the autoencoder encodes them as a sequence of points in its latent space. Then the sequences of latent attributes are smoothed through an optimization procedure.  The goal of the optimizer is to change the latent attributes as little as possible while at the same time enforcing the second-order temporal consistency indicator. Once each attribute has been corrected for temporal consistency, the resulting sequence of modified latent points is decoded back into 2D segmentation maps that are now temporally consistent.

A well-known class of methods to solve strictly-constrained optimization problems, applicable in our case, are those involving penalty functions such as barrier functions~\cite{sun_penalty_2006}. However, in practice, such functions require careful selection of parameters, and optimal parameters even depend on the degree of constraint violation~\cite{tessema_self_2006}. Thus, we rather opted to formulate the problem as a Lagrangian dual relaxation~\cite{ahuja_network_1993,rush_tutorial_2012}.

Given a sequence of attribute values in the latent space $s_a$, the goal is to find a new sequence of latent attribute  $s^*_a$ that minimizes the difference with $s_a$, under the strict constraint that $s^*_a$ satisfies the temporal Laplacian consistency indicator. Leveraging the notation defined in \cref{eq:temporal_indicator}, we can formulate the problem as:
\begin{equation}
\begin{aligned}
\label{eq:temporal_reg_problem}
& \arg\min_{s'_a} & & \| s_a - s'_a \|^{2}_{2}
\\ & \text{subject to} & & \mathbbm{1}(s'_a, t) = 0 \;\;\; \forall{t} \in \{0,1,\dots,T\} \quad 
\end{aligned}
\end{equation}
where $T$ is the total length of the sequence.

This formulation can be solved with a Lagrangian optimizer:
\begin{equation}
\arg\min_{s'_a} \underbrace{\| s_a - s'_a \|^{2}_{2}}_{f_a}+ \lambda_a \underbrace{\mathbbm{1}(s'_a, t)}_{g_a}
\end{equation}
where $f_a + \lambda_a g_a$ is the usual Lagrangian function. To solve this equation, the constraint has to be relaxed from the indicator function to a differentiable penalty function that gradually penalizes a frame the more it violates the constraint. The approximation of the Laplacian introduced in \cref{eq:local_linearity} is such a differentiable formulation, thus the relaxation $g'_a$ of the constraint $g_a$ can simply be the norm of the second-order derivative:
\begin{equation}
\label{eq:local_linearity_loss}
g'_a = \| s_{a,t+1} + s_{a,t-1} - 2s_{a,t} \|^{2}_{2}
\end{equation}
While $\lambda_a$ is often considered a hyperparameter, it is usually set to a constant value.  In our case, we want it to be as large as possible to force $g_a$ to be strictly $0$ but low enough such that $f_a$ still has an influence.  We do this by reformulating the problem as a dual Lagrangian relaxation min-max objective:
\begin{equation}
\arg \max_{\lambda_a} \min_{s'_a} f_a + \lambda_a g'_a
\end{equation}
where $\lambda_a$ is optimized like any other attribute variable $s'_a$.  Given the min-max formulation, $\lambda_a$ and $s'_a$ can be optimized via a subgradient iterative optimizer by alternatively optimizing both variables via a gradient ascent and gradient descent algorithm~\cite{rush_tutorial_2012}. However, because of computational efficiency concerns, we cannot maximize the threshold $\lambda_a$ using gradient ascent. This is because each inner optimization loop of $s'_a$ for a given $\lambda_a$ takes an average of 123 milliseconds. Therefore, we have to limit as much as possible the number of trials needed to find a suitable $\lambda_a$. We achieve this in practice by updating $\lambda_a$ through a binary search instead of gradient ascent, evaluating the temporal consistency indicators on each update to make sure the new value of $\lambda_a$ is large enough to strictly enforce the original constraint $g_a = 0$. Using this method, we are able to find $\lambda_a$ in only 5 updates, and optimizing a full sequence of 30 to 50 frames thus takes 6.4 seconds in total on average (because of multiple inner optimization loops and multiple attributes to optimize). All the times mentioned above were measured on a computer with an Intel Xeon E5-1620 v4 processor and a 12GB Nvidia Titan Xp graphics card.

For residual dimensions in the latent space, no empirical thresholds $\tau$ can be computed since they do not correlate to interpretable cardiac shape attributes. Consequently, no strict constraint $g_a$ exists, and we cannot determine an optimal $\lambda$. In that case, $\lambda$ becomes a simple hyperparameter that we determined empirically to be 50.

\begin{figure}[tp]
    \begin{subfigure}[b]{\columnwidth}
        \includegraphics[width=\textwidth]{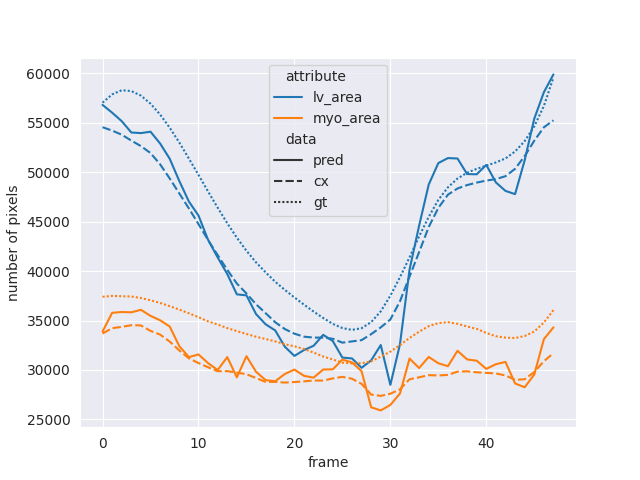}
        \caption{Corrected LV and MYO areas (I)}
    \end{subfigure}
    \begin{subfigure}[b]{\columnwidth}
        \includegraphics[width=\textwidth]{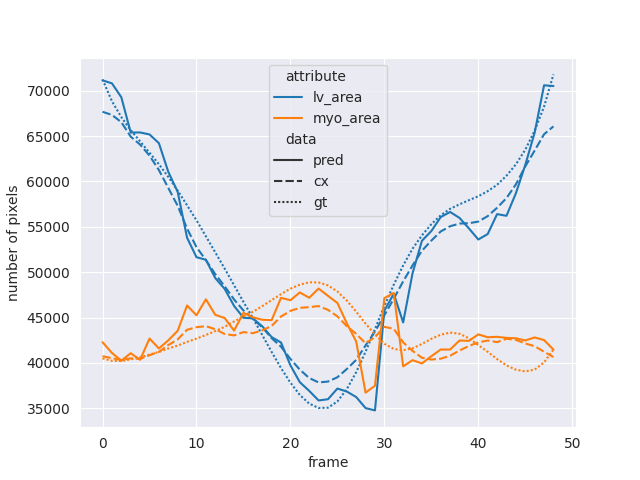}
        \caption{Corrected LV and MYO areas (II)}
    \end{subfigure}
    \begin{subfigure}[b]{\columnwidth}
        \includegraphics[width=\textwidth]{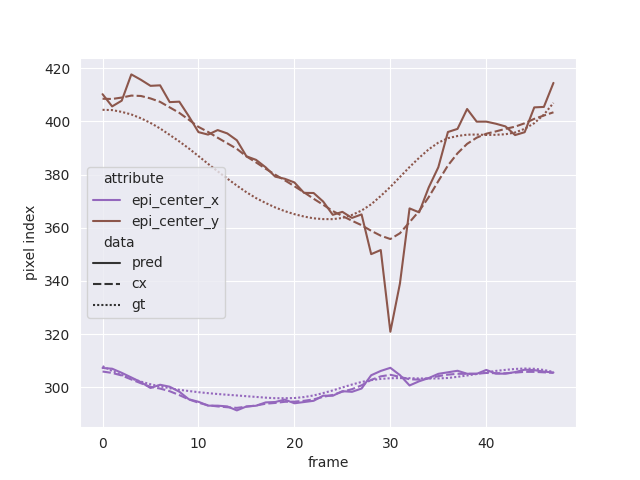}
        \caption{Corrected coordinates of the EPI's center of mass}
    \end{subfigure}
    \caption{Temporal inconsistencies, as the evolution of attributes w.r.t. time, before and after our temporal regularization method. The smooth dotted lines (\textit{gt}) correspond to attributes of the ground truth segmentations. The chaotic solid lines (\textit{pred}) represent the attributes extracted from a SOTA segmentation method's predictions. Finally, the smooth dashed lines (\textit{cx}) illustrate the attributes from the predictions corrected by temporal regularization.}
    \label{fig:temporal_regularization_plots}
\end{figure}

\section{Experiment Setup and Results}
\label{sec:setup_and_results}

\subsection{Dataset, Autoencoders and Segmentation Methods}
\label{sec:setup}

\subsubsection{CAMUS Full Cycle}
\label{sec:camus_full_cycle}
To test our method, we rely on the CAMUS dataset~\cite{leclerc_deep_2019}, presented in \cref{sec:echo_segmentation}. However, we complement the original dataset by providing brand-new manual segmentations of the LV and MYO classes in A4C sequences across the complete cardiac cycle for 98 of the 500 patients in the dataset. The segmentations were made by an experienced echocardiography specialist, and afterwards validated by a cardiologist. The sequences were also inspected to make sure there are no acquisition artifacts that would impact the temporal dynamics of the segmentation, e.g. foreshortening of the ventricle's apex. Just like in the initial release of the dataset, the segmentation protocol relies on identifying control points along the endocardium and epicardium, from which the pixel-wise boundaries are derived. To segment the whole cardiac cycle, our expert manually segmented eight frames in each sequence: both ED frames, the ES frame and 5 intermediate frames. Afterwards, they used spline interpolation on the manually segmented frames, i.e. the control points, to interpolate segmentations for the remaining frames. The interpolated contours were then checked manually and corrected if needed.

\subsubsection{Cardiac Shape Autoencoders}
\label{sec:cardiac_ar-vae_details}
In order to train both the baseline VAE and the cardiac AR-VAE, we used the ground truth segmentations from the CAMUS dataset~\cite{leclerc_deep_2019}, resized to 256$\times$256. The architecture used is most similar to the VAE network presented in Painchaud \etal~\cite{painchaud_cardiac_2020}, with added batch normalization after each non-linear activation layer. \Cref{tab:cardiac_ar-vae} describes in detail the architectures of the cardiac shape encoder and decoder, implemented in PyTorch v1.7.1~\cite{paszke_pytorch_2019}.

\begin{table}[tp]
    \centering
	\caption{Details of the VAE and cardiac AR-VAE's architecture. Both the encoder and the decoder blocks are repeated 4 times to iteratively downsample and upsample the images, respectively. The encoder and decoder projections map between the convolutional layers and the vectorized latent space. The first encoder block ($b_1$) outputs 48 filters, with each consecutive block doubling the number of filters. The numbers of filters in each decoder block mirror their corresponding encoder block. $b$:~block index, $o$:~output filters, $k$:~kernel size, $s$:~stride, $a$:~activation}
	\smallskip
	\begin{tabular*}{0.95\columnwidth}
        {@{} @{\extracolsep{\fill}} l l @{}}
		\toprule
		\multirow{2}*{Encoder Block ($\times$4)} & Conv o\textsubscript{b}=o\textsubscript{b-1}*2, k=3, s=2, a=ELU + BatchNorm \\
		& Conv k=3, s=1, a=ELU + BatchNorm \\
		\midrule
		\multirow{2}*{Encoder Projection} & Conv o=48 k=3, s=2, a=ELU + BatchNorm \\
		& 2 $\times$ Linear ($\mu$, $\sigma$ heads) o=16 \\
		\midrule \midrule
		Decoder Projection & Linear o=3072 \\
		\midrule
		\multirow{2}*{Decoder Block ($\times$4)} & ConvTrans k=2, s=2, a=ELU + BatchNorm \\
		& Conv k=3, s=1, a=ELU + BatchNorm \\
		\midrule
		\multirow{2}*{Decoder Output} & ConvTrans o=48, k=2, s=2, a=ELU + BatchNorm \\
		& Conv o=2, k=3, s=1, a=sigmoid \\
		\bottomrule
	\end{tabular*}
\label{tab:cardiac_ar-vae}
\end{table}

As for the hyperparameters, we performed a grid search over the parameters either with the most impact or most specific to our task: learning rate ($\eta$), weight decay ($\lambda$), number of latent dimensions ($d$), attribute regularization strength ($\gamma$) and spread of the posterior distribution ($\delta$). For the baseline VAE, the optimal hyperparameters were $\eta$~=~0.0005, $\lambda$~=~0.001, $d$~=~16 (since $\gamma$ and $\delta$ do not apply). In the case of the AR-VAE, the best hyperparameters were $\eta$~=~0.0005, $\lambda$~=~0.0005, $d$~=~16, $\gamma$~=~0.1, $\delta$~=~16. Both models were trained using Adam~\cite{kingma_adam_2015} with a batch size of 128 for 200 epochs.

\subsubsection{Segmentation methods}
\label{sec:segmentation_methods}
Since our proposed solution is a post-processing method that does not perform segmentation on its own, we have to rely on previous SOTA segmentation methods as a backbone. To provide a comprehensive analysis of how our method performs given input segmentations with varying degrees of accuracy and consistency, we post-processed the segmentations predicted by some generic segmentation architectures commonly used in echocardiography segmentation, i.e. DeepLabv3, ENet and UNet, as well as some more domain-specific methods, i.e. LUNet and CLAS. All these methods were trained on the CAMUS dataset. CLAS made use of more input US images, using intermediary frames between ED and ES, but all methods had access to the same 1800 2D segmentation maps for reference.

\begin{figure*}[tp]
    \begin{subfigure}[b]{\columnwidth}
        \includegraphics[width=\textwidth]{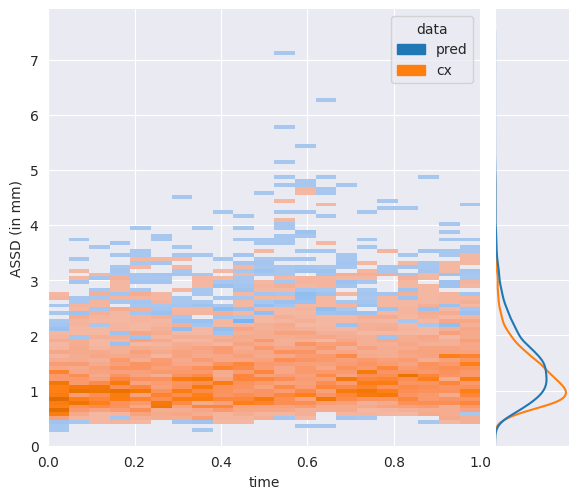}
        \caption{LV Average Symmetric Surface Distance (ASSD) w.r.t. time}
    \end{subfigure}
    \hfill
    \begin{subfigure}[b]{\columnwidth}
        \includegraphics[width=\textwidth]{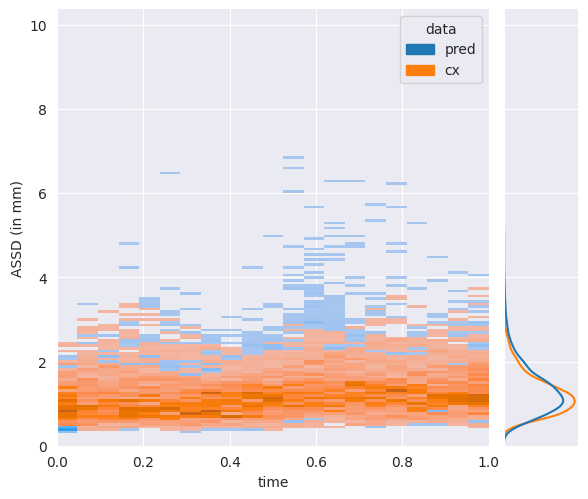}
        \caption{EPI ASSD w.r.t. time}
    \end{subfigure}
    \begin{subfigure}[b]{\columnwidth}
        \includegraphics[width=\textwidth]{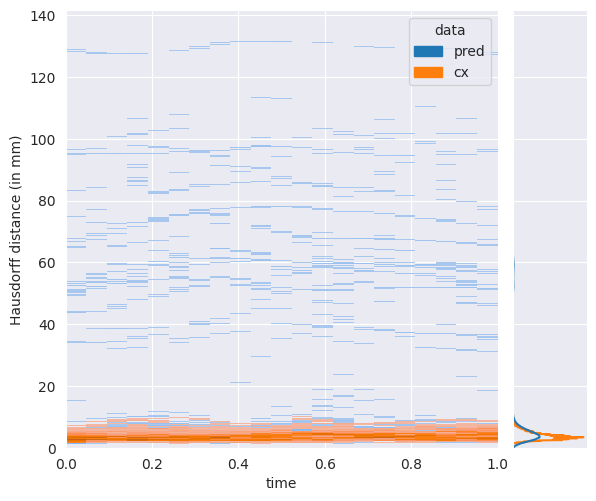}
        \caption{LV Hausdorff Distance (HD) w.r.t. time}
    \end{subfigure}
    \hfill
    \begin{subfigure}[b]{\columnwidth}
        \includegraphics[width=\textwidth]{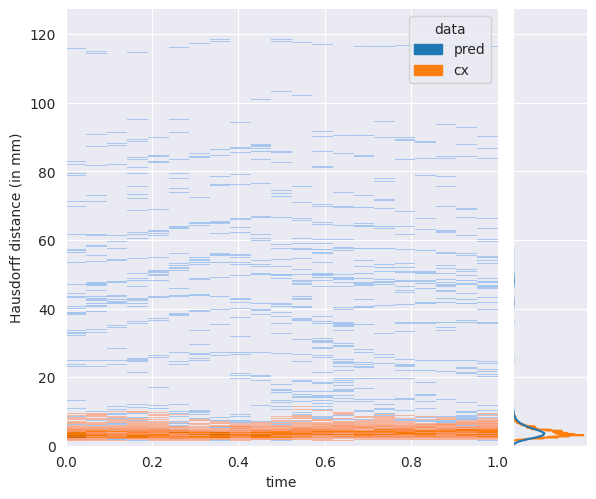}
        \caption{EPI HD w.r.t. time}
    \end{subfigure}
    \caption{Distribution of the segmentations' accuracy metrics w.r.t. time. For every plot, the x-axis represents the frame index normalized between [0,1], since the sequences contain varying numbers of frames in total, while the y-axis represents the metric's value. For these figures, the SOTA segmentation method was the UNet from Leclerc \etal~\cite{leclerc_deep_2019} (\emph{pred}) while our post-processed results are dubbed \emph{cx}.}
    \label{fig:temporal_regularization_accuracy_wrt_time}
\end{figure*}

An important caveat concerning the results reported in \cref{tab:temporal_regularization_accuracy_ablation_study,tab:temporal_regularization_clinical_ablation_study,tab:temporal_regularization_consistency_ablation_study} and \cref{fig:temporal_regularization_accuracy_wrt_time} is that they should not be compared to other results previously reported on CAMUS. This is because the 98 full cycles come from patients already part of CAMUS, including some from its training set. Since we could not train ourselves some of the benchmark methods, i.e. CLAS, we could not select the data the methods were trained on to exclude the ED and ES frames from the patients that overlap the 98 full-cycle testing sequences and CAMUS training set. However, we would argue that given the focus of our paper on the temporal consistency of the segmentation methods relative to one another rather than on the absolute accuracy of the ED and ES segmentations, the overlap does not detract from our conclusions. The priority is to ensure that the methods are compared fairly to one another instead of previous benchmarks.

\subsection{Experimental Results}
\label{sec:results}
We evaluate how well we enforce temporal consistency by post-processing results from the various SOTA segmentation methods mentioned in the previous section. We also compare ourselves to a baseline post-processing and by performing an ablation study of our pipeline in \cref{tab:temporal_regularization_accuracy_ablation_study,tab:temporal_regularization_clinical_ablation_study,tab:temporal_regularization_consistency_ablation_study}. In these tables, the \emph{Original} column corresponds to the predictions straight from the SOTA segmentation methods, with some standard morphological post-processing applied on each 2D frame independently, e.g. filling holes, removing blobs, etc. The \emph{Gaussian filter} column corresponds to the baseline temporal post-processing, where we apply a Gaussian filter along the temporal dimension on the segmentations, with a $\sigma$ empirically chosen to be $\frac{1}{20}$th of the number of frames in the sequence.

For our ablation study, we decided to analyse whether autoencoders by themselves have enough of a denoising effect to smooth out temporal inconsistencies. Thus, the \emph{VAE} column reports the results from simply reconstructing the segmentations, frame by frame, using a vanilla variational autoencoder. Next, the first of the \emph{Cardiac AR-VAE} columns does the same, but using our specialized cardiac shape autoencoder instead of a vanilla VAE. The next column, \emph{Temp. reg. attrs.}, comes from applying our temporal regularization operation only on our latent attributes, excluding the residual dimensions. The final column is our complete method, \emph{Temp. reg.}, and performs temporal regularization on both latent attributes and residual dimensions.

\begin{table*}[tp]
    \centering
	\caption{Temporal consistency of SOTA segmentation methods, with and without our temporal regularization. [Left] Number of sequences with at least one temporally inconsistent frame w.r.t its neighboring frames, out of a \textbf{total of 98 sequences}. [Right] Ratio between the Laplacian of an attribute at a given frame, as defined in \cref{eq:local_linearity}, and the maximum threshold above which the value for that attribute are considered temporal inconsistent. The ratio is averaged over all attributes and all frames, regardless of whether they are temporally consistent or not.
	}
	\smallskip
	\begin{tabular*}{0.8\textwidth}
        {@{} @{\extracolsep{\fill}} l c c c ccc @{}}
		\toprule
		\multirow{3}*{Methods} & \multicolumn{1}{c}{Original} & \multicolumn{1}{c}{Gaussian filter} & \multicolumn{1}{c}{VAE} & \multicolumn{3}{c}{Cardiac AR-VAE} \\
		\cmidrule(lr){2-2} \cmidrule(lr){3-3} \cmidrule(lr){4-4} \cmidrule(lr){5-7}
		& & & & \mcc{-} & \mcc{Temp. reg. attrs.} & \mcc{Temp. reg.} \\
		\midrule
        CLAS~\cite{wei_temporal-consistent_2020} & 68 / .211 & 41 / .143 & 56 / .199 & 54 / .212 & 1 / .083 & \textbf{0} / \textbf{.069} \\
        DeepLabv3~\cite{chen_encoder-decoder_2018,ouyang_video-based_2020} & 98 / .879 & 86 / .237 & 98 / .871 & 98 / .892 & 81 / .277 & \textbf{1} / \textbf{.125} \\
        ENet~\cite{paszke_enet_2016,painchaud_cardiac_2020} & 98 / .700 & 85 / .238 & 98 / .681 & 98 / .655 & 67 / .221 & \textbf{0} / \textbf{.114} \\
        LUNet~\cite{leclerc_lu-net_2020} & 98 / .594 & 79 / .214 & 98 / .591 & 98 / .641 & 61 / .220 & \textbf{0} / \textbf{.113} \\
        U-Net~\cite{ronneberger_u-net_2015,leclerc_deep_2019} & 98 / .562 & 83 / .218 & 98 / .558 & 98 / .584 & 53 / .198 & \textbf{0} / \textbf{.110} \\
		\bottomrule
	\end{tabular*}
\label{tab:temporal_regularization_consistency_ablation_study}
\end{table*}

\begin{table*}[tp]
    \centering
	\caption{Accuracy of SOTA segmentation methods, with and without our temporal regularization, on all frames of the full cycles from the CAMUS dataset. [Left] Average Dice score and [Right] Hausdorff distance (in mm).}
	\smallskip
	\begin{tabular*}{0.8\textwidth}
        {@{} @{\extracolsep{\fill}} l c c c ccc @{}}
		\toprule
		\multirow{3}*{Methods} & \multicolumn{1}{c}{Original} & \multicolumn{1}{c}{Gaussian filter} & \multicolumn{1}{c}{VAE} & \multicolumn{3}{c}{Cardiac AR-VAE} \\
		\cmidrule(lr){2-2} \cmidrule(lr){3-3} \cmidrule(lr){4-4} \cmidrule(lr){5-7}
		& & & & \mcc{-} & \mcc{Temp. reg. attrs.} & \mcc{Temp. reg.} \\
		\midrule
        CLAS~\cite{wei_temporal-consistent_2020} & \textbf{.953} / 4.4 & \textbf{.953} / 4.3 & \textbf{.953} / \textbf{4.0} & \textbf{.953} / \textbf{4.0} & \textbf{.953} / \textbf{4.0} & .952 / \textbf{4.0} \\
        DeepLabv3~\cite{chen_encoder-decoder_2018,ouyang_video-based_2020} & .946 / 4.8 & \textbf{.951} / 4.4 & .945 / 4.8 & .945 / 4.8 & .949 / 4.4 & \textbf{.951} / \textbf{4.2} \\
        ENet~\cite{paszke_enet_2016,painchaud_cardiac_2020} & .943 / 5.1 & .946 / 4.8 & .943 / 4.9 & .944 / 4.8 & .947 / 4.5 & \textbf{.949} / \textbf{4.3} \\
        LUNet~\cite{leclerc_lu-net_2020} & .947 / 4.6 & .951 / 4.3 & .947 / 4.6 & .947 / 4.6 & .950 / 4.3 & \textbf{.952} / \textbf{4.1} \\
        U-Net~\cite{ronneberger_u-net_2015,leclerc_deep_2019} & .951 / 4.3 & .954 / 4.0 & .951 / 4.3 & .950 / 4.3 & .953 / 4.1 & \textbf{.955} / \textbf{3.9} \\
		\bottomrule
	\end{tabular*}
\label{tab:temporal_regularization_accuracy_ablation_study}
\end{table*}

\begin{table*}[tp]
    \centering
	\caption{Clinical metrics of SOTA segmentation methods, with and without our temporal regularization. [Left] MAE on the 2D ejection fraction (EF) computed from A4C segmentations. [Right] Number of frames with anatomical errors, as defined in~\cite{painchaud_cardiac_2020}, out of a \textbf{total of 4531 frames} from all frames across all sequences.}
	\smallskip
	\begin{tabular*}{0.8\textwidth}
        {@{} @{\extracolsep{\fill}} l c c c ccc @{}}
		\toprule
		\multirow{3}*{Methods} & \multicolumn{1}{c}{Original} & \multicolumn{1}{c}{Gaussian filter} & \multicolumn{1}{c}{VAE} & \multicolumn{3}{c}{Cardiac AR-VAE} \\
		\cmidrule(lr){2-2} \cmidrule(lr){3-3} \cmidrule(lr){4-4} \cmidrule(lr){5-7}
		& & & & \mcc{-} & \mcc{Temp. reg. attrs.} & \mcc{Temp. reg.} \\
		\midrule
        CLAS~\cite{wei_temporal-consistent_2020} & 4.2 / 6 & 4.6 / 2 & 4.3 / 1 & 4.1 / 1 & 4.1 / \textbf{0} & \textbf{4.0} / \textbf{0} \\
        DeepLabv3~\cite{chen_encoder-decoder_2018,ouyang_video-based_2020} & 2.8 / 12 & \textbf{2.7} / 13 & 2.9 / 1 & \textbf{2.7} / 3 & \textbf{2.7} / 3 & 3.6 / \textbf{0} \\
        ENet~\cite{paszke_enet_2016,painchaud_cardiac_2020} & 3.2 / 11 & \textbf{2.9} / 20 & 3.4 / 5 & \textbf{2.9} / \textbf{0} & \textbf{2.9} / \textbf{0} & 3.3 / \textbf{0} \\
        LUNet~\cite{leclerc_lu-net_2020} & 3.1 / 11 & \textbf{2.7} / 8 & 3.1 / \textbf{0} & \textbf{2.7} / 1 & 2.8 / 1 & 3.6 / \textbf{0} \\
        U-Net~\cite{ronneberger_u-net_2015,leclerc_deep_2019} & \textbf{2.7} / 1 & 2.9 / 6 & 2.9 / 6 & 2.8 / 3 & 3.0 / \textbf{0} & 3.8 / \textbf{0} \\
		\bottomrule
	\end{tabular*}
\label{tab:temporal_regularization_clinical_ablation_study}
\end{table*}

Finally, to evaluate how temporal regularization would affect abnormally rapid but temporally consistent motions, we manually inspected all 98 sequences to determine if any of them exhibited cases of “extreme” temporal dynamics and/or known pathological signatures. We present an in-depth analysis of a representative case of rapid motion in \cref{fig:temporal_regularization_limitations}.

In the following sections, we discuss in details the results from \cref{tab:temporal_regularization_accuracy_ablation_study,tab:temporal_regularization_clinical_ablation_study,tab:temporal_regularization_consistency_ablation_study} and the example from \cref{fig:temporal_regularization_limitations}, in the context of what they mean for the clinical interpretability of the segmentations.

\subsubsection{Enforced temporal and anatomical plausibility of the segmentations}
\label{sec:results:temporal_anatomical_plausibility}
\Cref{tab:temporal_regularization_consistency_ablation_study} details each method's temporal consistency by providing i) the numbers of sequences with at least one temporally inconsistent frame, i.e. frames where at least one attribute fails the temporal consistency indicator defined by \cref{eq:temporal_indicator}, out of the 98 sequences in the dataset, and ii) the average ratio between the Laplacian of an attribute and its threshold. Note that a large ratio means that, on average, inconsistencies are far above the associated threshold whereas a value close to 1 would mean the temporal dynamics of the attributes are near the inconsistency threshold. We see that even SOTA 2D segmentation methods fail to be temporally consistent, leaving inconsistencies in all the sequences, and only the 3D method, CLAS, performs better. Though even then, it is inconsistent in nearly 70\% of sequences. The baseline Gaussian filter helps but is not reliable, failing to correct the majority of inconsistencies. From the \emph{VAE} and \emph{Cardiac AR-VAE} columns, we also confirm the intuitive notion that 2D autoencoders are not capable of denoising temporal artifacts. The consistency starts to improve once we begin to apply our temporal regularization in \emph{Temp. reg. attrs.}, though mostly for CLAS since the temporal artifacts were less severe to begin with. However, we need to include residual dimensions, in \emph{Temp. reg.}, to reliably obtain temporal consistency, where we only fail for one especially degenerated sequence from DeepLabv3.

While our method enforces sequences to be temporally plausible, it is interesting to note how this affects the anatomical plausibility of each frame individually. In \cref{tab:temporal_regularization_clinical_ablation_study}, the results on the right in each column count the anatomically inconsistent frames from the 4,531 frames in total in the dataset, following the anatomical criteria defined for long-axis cardiac shapes by Painchaud \etal~\cite{painchaud_cardiac_2020}. Here, we see an obvious limitation of the post-processing baseline, which worsens the anatomical plausibility for the DeepLabv3, ENet and U-Net methods. This can be explained by the Gaussian filter having no global semantic understanding of the content of the image, and thus introducing anatomically inconsistent artifacts at the frontiers between classes. In contrast, all the results from autoencoders reduce the numbers of anatomically erroneous frames because of their implicit 2D denoising capabilities. Empirically, we also noticed that the most egregious anatomical errors tend to also lead to temporal inconsistencies. This explains why our full method, which modifies the temporal inconsistent frames the most, also implicitly corrects their anatomical errors.

\subsubsection{Improvements in segmentation accuracy derived from temporal consistency}
\label{sec:results:accuracy}
Intuitively, we can hypothesize that improving the temporal consistency of segmentations would improve their accuracy. \Cref{tab:temporal_regularization_accuracy_ablation_study} supports this hypothesis across the board. We can see gradual, if slight, improvements in segmentation accuracy by first reconstructing the segmentations using autoencoders (\emph{VAE}, \emph{Cardiac AR-VAE}), and then applying our method on some or all dimensions in the latent space (\emph{Temp. reg. attrs.}, \emph{Temp. reg.}). Because the predictions are sometimes markedly off from the ground truth, e.g. see curves in \cref{fig:temporal_regularization_plots} (a) and (c), improving the temporal consistency won't necessarily bring the predicted segmentations much closer to the ground truth on average. This explains why the post-processing does not improve the Dice score coefficient (DSC) significantly. Where the impact is noticeable is on the Hausdorff distance (HD), which our post-processing improves by 0.5mm on average across all methods. Like explained in the previous section, this is because we can rely on the temporal context of a frame to correct it more heavily when it is inconsistent, and egregious errors in a frame are likely to cause temporal inconsistencies. The behavior of our method with respect to outliers is further illustrated in \cref{fig:temporal_regularization_accuracy_wrt_time}, where we plot the distributions of the ASSD and HD with respect to time for both the SOTA predictions (in blue) and the corrections from our post-processing (in orange). The predictions contain a significant number of outliers, especially for the HD in (c) and (d), with outliers of as much as 135mm. These outliers, being the result of aberrations localized in space and time, completely disappear in our post-processed results, leading the HD values to be concentrated in the lower end of the spectrum. As for the ASSD in (a) and (b), the improvements are more modest, since the metric has a global scope and therefore is less sensitive to localized aberrations than to an overall misalignment from the ground truth, which our temporal consistency cannot fix. However, even then we can still observe less outliers.

\subsubsection{Similar behavior to smooth reference segmentations on rapid local motions}
\begin{figure}[tp]
    \centering{
    \begin{subfigure}[b]{.53\columnwidth}
        \includegraphics[width=\textwidth]{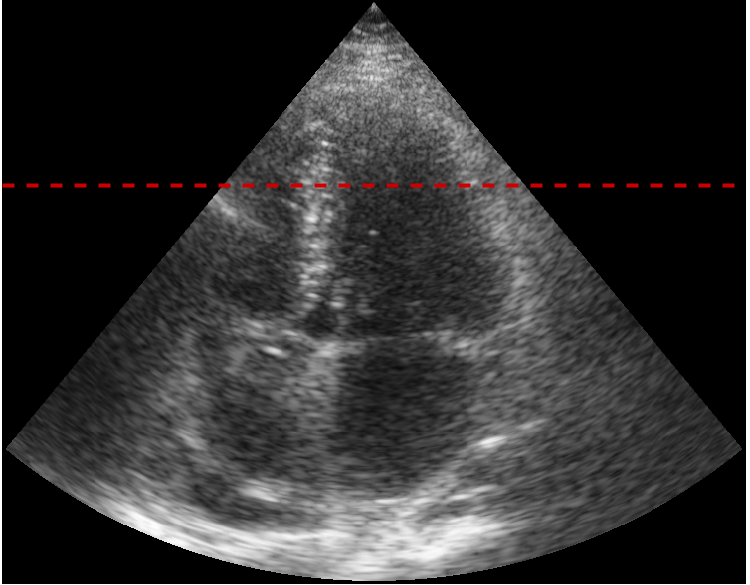}
        \caption{Plane of the cut for figures b) through e)}
    \end{subfigure}
    }
    \vfill
    \begin{subfigure}[b]{.49\columnwidth}
        \includegraphics[width=\textwidth]{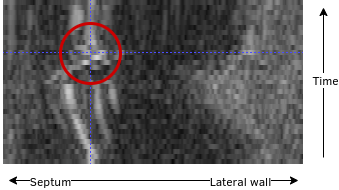}
        \caption{Echocardiography}
    \end{subfigure}
    \hfill
    \begin{subfigure}[b]{.49\columnwidth}
        \includegraphics[width=\textwidth]{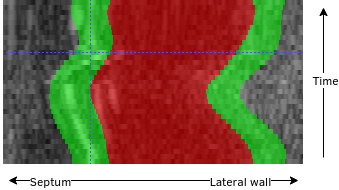}
        \caption{Ground truth segmentation}
    \end{subfigure}
    \begin{subfigure}[b]{.49\columnwidth}
        \includegraphics[width=\textwidth]{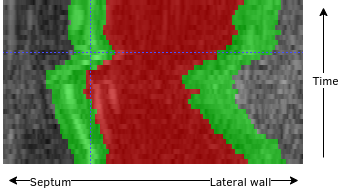}
        \caption{U-Net prediction}
    \end{subfigure}
    \hfill
    \begin{subfigure}[b]{.49\columnwidth}
        \includegraphics[width=\textwidth]{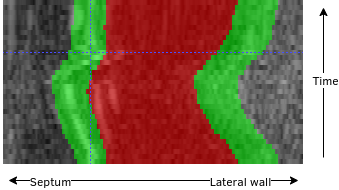}
        \caption{U-Net after temp. reg.}
    \end{subfigure}
    \caption{Impact of temporal regularization on a patient with rapid inward/outward radial motion near the apex during early diastole. Each image shows the same radial cut of the image across the whole sequence (a), with figures c), d) and e) overlaying different segmentations on top of the echocardiographic image (b). Figure b) marks where the rapid movement occurs. The motion was captured in the U-Net prediction (d), but was lost in both the smooth ground truth (c) and the temporal regularization of the U-Net (e).}
    \label{fig:temporal_regularization_limitations}
\end{figure}

When manually inspecting the echocardiographic sequences, the most prominent pattern we observed (on 5 to 10 subjects) consists of a fast inward/outward radial motion near the apex during early diastole, i.e. post-systolic abnormal motion. As visible in \cref{fig:temporal_regularization_limitations}, which shows the most extreme of these identified cases, the rapid motion was adequately captured in the U-Net prediction. Still, the U-Net prediction was temporally noisy around the rapid motion and in the vicinity of the lateral wall and the apex. While temporal regularization denoised the segmentation, it also smoothed over the rapid motion, but no more than the ground truth segmentations themselves. The temporal regularization behaves similarly on the other patients with rapid motion, i.e. it is closer to the interpolated ground truth segmentations than the original predictions.

\subsubsection{No significant deterioration of traditional scalar clinical metrics}
\label{sec:results:clinical}
The metric commonly used to measure the clinical potential of a segmentation algorithm is the ejection fraction (EF) between the ED and ES instants of the cardiac cycle. While the best estimation of the EF is normally computed using Simpson's biplane method when two orthogonal views are available~\cite{folland_assessment_1979}, the 2D LV area can be used as an acceptable substitute if only one view is available~\cite{ouyang_video-based_2020}, like with CAMUS Full Cycle. However, using the 2D area favors exact segmentations of individual frames over consistent spatio-temporal segmentations, and might explain why CLAS does not outperform other methods like reported in the original paper. Thus, \cref{tab:temporal_regularization_clinical_ablation_study} reports the MAE on the ejection fraction (left number in each column). On average, the MAE is slightly worse after post-processing than before, increasing by 0.5\%. The worst case is the U-Net, where the MAE increases from 2.7 to 3.8. Although this might appear substantial at first glance, EF measurements are known to suffer from bad reproducibility, with absolute variations around 3 \%~\cite{thorstensen_reproducibility_2010}.

As for an explanation for this degradation, it can be linked to temporal regularization smoothing spikes in the attributes at ED and ES, which especially affects the ED instants at the extremities of the sequences. Even though our method uses edge-padding on the sequences to avoid cyclical assumptions on the data, the degradation of the segmentations is more pronounced at ED than at ES. Improvements to how we handle the end points of the sequence could potentially alleviate the problem. However, because temporal regularization naturally affects peaks in the signal the most, a slight baseline degradation of ED/ES segmentations, and therefore EF, might be inevitable. Still, we want to point out that even if the error on the EF is not improved, the average absolute difference between the EF MAE after our temporal regularization and the EF MAE on the original predictions is 0.5\%, which is well within the reported intra-observer variability on CAMUS~\cite{leclerc_deep_2019} and similar to variations introduced by reproducibility issues already assumed by clinicians, and therefore not a cause for serious concerns.

We would also argue that since our main objective is a consistent segmentation across the full cycle, and not specifically the best possible segmentations for the ED and ES instants alone, we are hitting the limits of the EF as an all-encompassing metric describing the cardiac function. As such, while the reported EF errors suggest similar clinical usability of the predictions before and after our temporal regularization, \cref{fig:temporal_regularization_plots} clearly illustrates that variations over time of our post-processed predictions (dashed lines) are much closer to the ground truth (dotted lines) than the original SOTA predictions (solid lines). In the clinical literature, studies have argued for years about the relevance of descriptors that are more complex than the EF and can describe the cardiac function across the whole cycle~\cite{cikes_beyond_2016}, e.g. motion or deformation, even computed globally along the longitudinal direction~\cite{reisner_global_2004}. We believe that the temporal consistency we introduce can serve as a catalyst to the widespread adoption of these more complex descriptors of the cardiac function by making it possible to automatically compute them, accurately and reliably, from segmentations of echocardiographic sequences.

\section{Conclusion}
\label{sec:conclusion}
We proposed a post-processing pipeline to enforce temporal consistency in 2D+time echocardiography segmentation. The temporal consistency is enforced as a constrained regularization on the curves with respect to time of seven clinically relevant attributes that describe 2D long-axis cardiac shapes. We also relied on these attributes to learn an interpretable cardiac shape autoencoder, which is used in the post-processing pipeline to coherently correct the temporal inconsistencies left behind by segmentation methods. We tested our post-processing on five SOTA segmentation methods, both generic and specialized for echocardiography segmentation, and showed systematic improvements on segmentation accuracy metrics, i.e. Dice score coefficient and Hausdorff distance, on top of the enforced temporal consistency. Another advantage of our method is that it does not require 2D+time annotated data, since our cardiac shape autoencoder is trained on 2D images. The temporal consistency derives entirely from how we make use of the latent space's interpretability.

To properly evaluate our post-processing on the full cycle, we also provided new annotations for the A4C view for 98 out of the 500 patients in the CAMUS dataset.

Finally, from a clinical perspective, automatic tools are currently mainly used to segment the ED and ES instants and to provide scalar clinical indices based on these, like the EF. However, many other indices frequently used, like velocities, strain, or strain rate, require temporally consistent data and can serve to examine the whole cardiac cycle. Thus, temporally consistent segmentation is a necessary step towards automatic image analysis tools that can help to identify and characterize a wider range of cardiovascular diseases.



\section*{Acknowledgment}

The authors thank i) Eric Saloux (CHU Caen, France) for his guidance on the manual segmentations of temporal sequences, ii) Compute Canada for providing access to computing resources, and iii) Wei \etal~for supplying the predictions of their CLAS method on the full cycle US sequences.


\bibliographystyle{IEEEtran}
\bibliography{TMI-2021-2030}





\end{document}